\documentclass{article}

\usepackage{microtype}
\usepackage{graphicx}
\usepackage{subfigure}
\usepackage{booktabs} 

\usepackage{hyperref}

\usepackage[accepted]{icml2023}

\usepackage{amsmath}
\usepackage{amssymb}
\usepackage{mathtools}
\usepackage{amsthm}

\usepackage[capitalize,noabbrev]{cleveref}

\usepackage{amsmath,amsfonts,bm}

\def\eqref#1{equation~\ref{#1}}

\def\1{\bm{1}}

\def\vs{{\bm{s}}}

\def\mC{{\bm{C}}}

\def\mS{{\bm{S}}}

\def\mW{{\bm{W}}}

\DeclareMathAlphabet{\mathsfit}{\encodingdefault}{\sfdefault}{m}{sl}
\SetMathAlphabet{\mathsfit}{bold}{\encodingdefault}{\sfdefault}{bx}{n}

\def\sD{{\mathbb{D}}}

\def\sU{{\mathbb{U}}}

\newcommand{\Lagr}{\mathcal{L}}
\newcommand\norm[1]{\left\lVert#1\right\rVert}

\begin{document}

\twocolumn[
\icmltitle{Enhancing the Stability of LLM-based Speech Generation Systems through Self-Supervised Representations}

\begin{icmlauthorlist}
\icmlauthor{Álvaro Martín-Cortinas}{amzn_before}
\icmlauthor{Daniel Sáez-Trigueros}{amzn}
\icmlauthor{Iván Vallés-Pérez}{amzn}
\icmlauthor{Biel Tura-Vecino}{amzn}
\icmlauthor{Piotr Biliński}{amzn,uow}
\icmlauthor{Mateusz Lajszczak}{amzn}
\icmlauthor{Grzegorz Beringer}{amzn}
\icmlauthor{Roberto Barra-Chicote}{amzn}
\icmlauthor{Jaime Lorenzo-Trueba}{amzn_before}
\end{icmlauthorlist}

\icmlaffiliation{amzn}{Amazon AGI}
\icmlaffiliation{amzn_before}{Work conducted at Amazon AGI but author is no longer affiliated}
\icmlaffiliation{uow}{University of Warsaw}
\icmlcorrespondingauthor{Daniel Sáez-Trigueros}{dsaez@amazon.com}
\icmlcorrespondingauthor{Roberto Barra-Chicote}{rchicote@amazon.com}

\vskip 0.3in
]

\printAffiliationsAndNotice{}

\begin{abstract}
Large Language Models (LLMs) are one of the most promising technologies for the next era of speech generation systems, due to their scalability and in-context learning capabilities. Nevertheless, they suffer from multiple stability issues at inference time, such as hallucinations, content skipping or speech repetitions. In this work, we introduce a new self-supervised Voice Conversion (VC) architecture which can be used to learn to encode transitory features, such as content, separately from stationary ones, such as speaker ID or recording conditions, creating speaker-disentangled representations. Using speaker-disentangled codes to train LLMs for text-to-speech (TTS) allows the LLM to generate the content and the style of the speech only from the text, similarly to humans, while the speaker identity is provided by the decoder of the VC model. Results show that LLMs trained over speaker-disentangled self-supervised representations provide an improvement of 4.7pp in speaker similarity over SOTA entangled representations, and a word error rate (WER) 5.4pp lower. Furthermore, they achieve higher naturalness than human recordings of the LibriTTS test-other dataset. Finally, we show that using explicit reference embedding negatively impacts intelligibility (stability), with WER increasing by 14pp compared to the model that only uses text to infer the style.
\end{abstract}

\section{Introduction} \label{sec:introduction}
The main goal of speech generation systems is to generate high-quality speech with a naturalness and expressiveness close to that of humans. Ideally, these models should be able to handle multiple speakers and multiple styles, even unseen ones. The arrival of Large Language Models (LLMs) to natural language processing~\citep{brown20-gpt3} proved that their in-context learning capabilities allow them to solve zero-shot tasks. Henceforth, there have been multiple attempts to leverage those capabilities into general Text-to-Speech (TTS) systems that are capable of generating any speaker with any style.

One of the first successful approaches to a multi-speaker and multi-style LLM-based TTS system is TorToiSe~\citep{betker23-tortoise}, where a vector-quantized variational auto-encoder (VQ-VAE) is trained to encode speech into discrete representations, an LLM is trained to predict VQ-VAE discrete codes directly from the text and a diffusion model followed by a vocoder generates the final speech from the discrete codes sampled from the LLM. Nevertheless, this model suffers from instabilities in the speaker ID and prosody due to an unstable sampling process in the LLM. As the speech codes are entangled in the VQ-VAE, the LLM is responsible for modelling the joint probability distribution of content, speaker, style, prosody, recording conditions and all the aspects related with the training data distribution. Furthermore, due to the entanglement of the codes, a reference embedding, computed by learning a compressed representation of the reference spectrogram, is passed to the LLM to determine the speaker identity and style the model should generate. This additional conditioning seems to introduce instabilities in the generation when the reference utterance is out-of-distribution and the reference embedding doesn't effectively represent the reference speech, which leads to CLIP-like models being used to rank outputs.

In a similar way, other architectures have been proposed to predict speech codes without the necessity of a reference embedding, such as VALL-E~\citep{wang23-valle} and VALL-E X~\citep{zhang23-vallex} which predict the codes of EnCodec~\citep{défossez22-encodec}. As no reference embedding is used, but the codes are entangled, the reference speaker information is passed to the model at inference time by conditioning the LLM with the transcription and speech codes of the reference. In practice, this results in a need for short references, because the reference length limits the maximum possible generated length. It also introduces a need for transcriptions, which are not always available or are prune to errors if they are computed with automatic speech recognition (ASR) systems, specially for uncommon languages.

Independently, and since the BERT~\citep{devlin19-bert} was published, masked language modeling (MLM) has been a recurrent pre-text task used in multiple self-supervised learning (SSL) models for speech. Two of the most important ones that use this pre-text task, according to the SUPERB benchmark~\citep{yang2021-superb}, are HuBERT~\citep{hsu21-hubert} and WavLM~\citep{chen22-wavlm}.

The masked language modeling task consists on predicting a randomly masked token from the input given the previous and following tokens, so that given a set of discrete tokens $\sU=\{u_1,\ldots,u_n\}$ and a randomly masked token $i$, the following likelihood is maximized:
\begin{equation}
    \mathcal{L}(\sU) = \sum_i P(u_i|u_1,\ldots,u_{i-1},u_{i+1},\ldots,u_{n}; \theta),
\end{equation}
where $\theta$ are the model's parameters. As the tokens of an MLM are already highly correlated, because they can be predicted from other tokens in the same sequence, the causal language modeling task performed in LLMs is expected to be simpler when predicting these tokens than when predicting tokens with no enforced correlation, such as the ones coming from audio codecs like EnCodec or HiFi-Codec~\citep{yang23-hificodec}.

\textbf{Contribution.} In this work, we present a new approach towards building LLM-based TTS systems by using speaker-disentangled and quantized self-supervised representations as targets for the LLM, which allow the LLM to focus on generating the non-speaker information such as content and style, improving the overall stability. This new approach is based on training a Self-Supervised Voice Conversion (SSVC) model that will learn speaker-disentangled and quantized self-supervised representations. To do so, WavLM will be used as encoder and contrastive learning will be leveraged to separate its hidden states into a speaker embedding\footnote{The speaker embedding is an internal representation learnt within the VC model and cannot be used for identification purposes.} and non-speaker features, which will be quantized to train an LLM with them. To the best of our knowledge, this is the first VC model that can be trained with non-parallel data and no annotations or pre-trained speaker embeddings, which facilitates its application to new datasets. We compare this model with the existing SOTA text-free VC model, FreeVC, which is also based on WavLM, and we show that our proposal outperforms it in intelligibility, in speaker similarity and in keeping the source utterance style.

Once SSVC has learned the non-speaker discrete representations, we train an LLM over these tokens, which as previously mentioned are expected to be predicted more easily than tokens from audio codecs. In this paper two versions of the LLM are trained, with and without reference embedding, to measure its impact in the content stability of the LLM. These models are abbreviated as LSSL-R and LSSL-NR, respectively. As the speaker is provided by SSVC, with this approach the speech can be generated by conditioning the LLM only with text, and no reference embedding or speech codes are needed like in TorToiSe and VALL-E. We compare the best LLM-based TTS model trained, LSSL-NR, with the current SOTA multi-speaker TTS system YourTTS, and we outperform it in intelligibility, naturalness and signal quality.

In summary, we make the following contributions:
\begin{itemize}
    \item We propose a new SOTA VC model, SSVC, based on self-supervised learning. This model can be trained only with audio, in contrast with previous models which required text, parallel data or pre-trained embeddings.
    \item We propose a new SOTA TTS model based on predicting the quantized and speaker-disentangled self-supervised features of SSVC. This TTS achieves a naturalness in its generation higher than human recordings in LibriTTS~\citep{zen19-libritts} test-other dataset.
    \item We prove the improvement in the stability over previous LLM-based speech generation inference techniques by using a new prompting technique based only on text, which is only possible due to predicting speaker-disentangled features.
\end{itemize}

\section{Related work}

\textbf{Voice conversion.} There are two types of VC systems~\citep{sisman20-vcoverview}: supervised (trained with parallel data) and unsupervised (trained with non-parallel data). Parallel data, recordings with the same content and different speaker ID, is expensive to acquire or generate~\citep{chen22-dataaug-nonparallelvc}, which makes it unfeasible for large amounts of recordings. Unsupervised approaches can be trained in a reconstruction setup, but disentangling speaker and non-speaker information is a challenge and there are multiple proposals in the literature: a bottleneck based on information theory by~\cite{qian19-autovc}, a variational autoencoder~\cite{saito18-vae}, generative adversarial networks (GAN) by~\cite{kaneko19-cycleganvc2, kameoka18-starganvc}, normalizing flows~\cite{merritt2022-textfree-nfvc} and self-supervised speech representations (S3Rs)~\cite{li22-freevc, lin21-fragmentvc, huang22-s3rvc}. In~\cite{huang22-s3rvc} it is also proved that quantizing S3Rs removes the speaker identity, although the prosody and the content might be also lost.

\textbf{Discrete audio codecs.} Discrete audio codecs allow to encode speech information into quantized representations, that can later be used to reconstruct the original waveform. In~\cite{zeghidour21-soundstream}, a residual vector quantization (RVQ) layer is proposed that quantizes the quantization error multiple times with consecutive codebooks to improve the quality of the reconstruction. In~\cite{défossez22-encodec}, perceptual loss based on a multi-scale STFT-based (MS-STFT) discriminator is added to improve the subjective quality of the codec. In~\cite{yang23-hificodec}, they propose the technique group-residual vector quantization (GRVQ), which consists on dividing in 2 the feature space and quantizing it separately to achieve more quality with less codebooks. All these systems learn entangled codes in contrast with our codec proposal. In this work we use EnCodec~\citep{défossez22-encodec} as a baseline. It is publicly available and has been used in SOTA TTS VALL-E.

\textbf{Speech generation based on discrete tokens.} Apart from the already mentioned VALL-E and VALL-E X, there are other speech generation systems based on LLMs that predict quantized tokens from audio codecs. In~\cite{borsos23-soundstorm}, MaskGIT~\citep{chang22-maskgit} is utilized to predict the quantized codes of SoundStream taking as input the semantic tokens of AudioLM~\citep{borsos23-audiolm}. In~\cite{kharitonov23-speartts}, the semantic tokens of w2v-BERT~\citep{chung2021-w2vbert} are mapped to the entangled speech codes of SoundStream. In both approaches, the speaker conditioning is performed like in VALL-E, prompting with the speech codes of a short reference of the desired speaker.

\section{Method}
\subsection{Problem formulation}
The process of creating an LLM-based TTS system can be separated in two sequential steps. First, given a dataset composed of $L$ utterances of raw speech $\sD_r = \{x_i(t)\}_{i=1}^L$, an audio codec is trained to reconstruct the speech using discrete representations: $\tilde{x}(t) = f^d_{\theta_2}(q[f^e_{\theta_1}(x(t))])$, where $q[\cdot]$ is the quantization algorithm, $f^e_{\theta_1}$ is the encoder function parametrized by $\theta_1$ and $f^d_{\theta_2}$ is the decoder function parametrized by $\theta_2$. In our particular case, the quantized representation should be disentangled in the encoder from the speaker information, which results in
\begin{equation}
    \mC(t), \vs = f^e_{\theta_1}(x(t)),\quad \tilde{x}(t) = f^d_{\theta_2}(q[\mC(t)], \vs),
\end{equation}
where $\mC(t)$ represents the non-stationary information of the speech, with one vector per time step, and $\vs$ represents the stationary information in the waveform, with one vector per waveform and including the global speaker ID information.

Once the audio codec is trained, it can be utilized to encode each element of the dataset as a sequence of discrete representations, which results in a new dataset for an LLM. In the particular case of this paper, the quantized representations predicted by the LLM are speaker independent, so the dataset used to train the LLM can be expressed as $\sD_q = \{q[\mC(t)] : \mC(t), \vs \in \{f^e_{\theta_1}(x(t)) : x(t) \in \sD\}\}$. With the LLM trained, the waveforms can be generated by converting the discrete tokens back to speech using the decoder of the audio codec.

Moreover, as the audio codec can disentangle stationary and non-stationary information, and the global speaker ID is part of the stationary information, the codec can be used as voice conversion system. To do so, given two waveforms, source and target, both must be encoded with $f^e_{\theta_1}$ and the speaker ID of the source can be converted to the target by doing $\tilde{x}_{converted}(t) = f^d_{\theta_2}(q[\mC_{source}(t)], \vs_{target})$.

\subsection{Self-supervised voice conversion model} \label{subsec:ssl-vc}
\textbf{Overview.} Disentangling speaker and non-speaker information using a completely self-supervised approach, without any type of labelled data, is a problem that, to the best of our knowledge, has not been addressed before in the literature. Therefore, a new architecture is proposed based on the SOTA self-supervised speech model WavLM~\citep{chen22-wavlm}.

\begin{figure*}[h]
  \centering
  \includegraphics[width=0.8\textwidth]{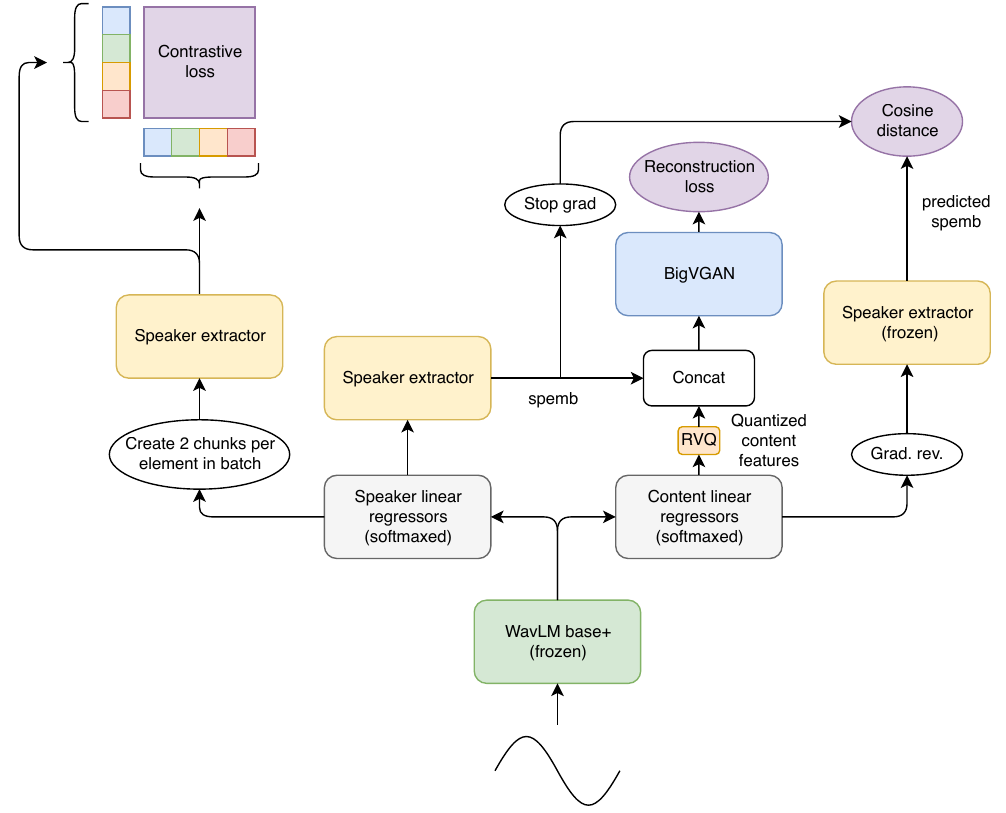}
  \caption{Self-supervised voice conversion architecture to learn speaker-disentangled S3R.}
  \label{fig:rvq-chameleon2}
\end{figure*}

A summary of the architecture is presented in Figure~\ref{fig:rvq-chameleon2}. It can be divided in three parts: learning the speaker features (left), learning the quantized non-speaker features (right) and learning to reconstruct the input (center). All of them are based on WavLM hidden states. In particular, to compute the speaker features the hidden states of all of WavLM's layers are averaged with learnable softmaxed-normalized speaker weights ($\text{softmax}(\mW_s)$). Then, to compute the non-speaker features an analogous procedure is followed, but the softmaxed-normalized non-speaker weights are constrained to $1 - \text{softmax}(\mW_s)$ to prevent one hidden state from providing speaker and non-speaker information at the same time.

\textbf{Speaker embedding.} The speaker embedding is learned using a contrastive loss following the work of~\cite{radford21-clip, elizalde22-clap}. The assumption behind this decision is that each element in the batch comes from a different speaker, which is reasonable when a large and varied dataset is being used to train the audio codec, as the speech time per speaker will be relatively small. Then, two chunks of the same element in the batch are considered positive samples, and two chunks of different elements in the batch are considered negative samples.

Given a utterance or a chunk, the speaker embedding is extracted from WavLM's hidden states by computing the speaker features with the softmaxed-normalized speaker weights, stacking a learnable embedding \texttt{CLS}~\citep{devlin19-bert} and passing the result through a transformer encoder which maps the speaker information into the \texttt{CLS}. That \texttt{CLS} embedding is then linearly projected to the desired dimensionality for the speaker embedding.

\textbf{Quantized non-speaker features.} As the speaker embedding is being learned minimizing the contrastive loss, the cosine similarity between speaker embeddings can be used as a metric of how similar are two speakers. Therefore, minimizing the absolute of the cosine similarity between embeddings computed using the transformer encoder on speaker and non-speaker features ensures there is no speaker information in the non-speaker features. Consequently, the gradient reversal layer~\citep{schnell21-emocat} is included so that the non-speaker linear regressors are trained to maximize the cosine distance instead of minimizing it. Then, the non-speaker features can be quantized using an RVQ layer to facilitate a high-quality reconstruction, following~\cite{défossez22-encodec} and~\cite{zeghidour21-soundstream}. The quantization is needed to obtain discrete tokens that can be predicted with the LLM.

\textbf{Reconstruction.} The reconstruction of the original waveform with high-quality requires having both speaker and non-speaker information. The non-speaker features are time-dependent, whereas the speaker embedding is computed on the whole utterance. Consequently, to reconstruct the input the speaker embedding is concatenated to the non-speaker features and the result is fed to the decoder, a BigVGAN~\cite{lee2023-bigvgan} model, to generate the final speech.

Taking all the different parts into account, the total loss of the generator during training can be mathematically described as shown in~\eqref{eq:total-generator-loss}:
\begin{multline} \label{eq:total-generator-loss}
    \Lagr_G = \lambda_0\Lagr_{BigVGAN} + \lambda_1\Lagr_{contrastive}(\mS) - \\ \lambda_2\left(1 - \left\vert\frac{\vs_s\cdot\vs_{ns}}{\norm{\vs_s}\norm{\vs_{ns}}}\right\vert\right) + \lambda_3\Lagr_{commitment},
\end{multline}
where $\mS$ is the square matrix with positive samples in the diagonal needed to compute the cross entropy loss, $\vs_s$ is the embedding computed with the speaker extractor and the speaker features, $\vs_{nc}$ is the embedding computed with the speaker extractor and the non-speaker features, $\Lagr_{commitment}$ is the commitment loss of the RVQ codebooks and $\lambda_0, \lambda_1, \lambda_2, \lambda_3$ are normalized hyperparameters to weight the contribution of each loss. The negative symbol before $\lambda_2$ comes from the gradient reversal layer, as only the parameters contributing to the content weights are being modified due to this loss.

\subsection{RVQ prediction}
In this work, the codes are being flattened by interleaving them and they are predicted autoregressively by all the LLMs, following~\cite{kharitonov23-speartts, borsos23-audiolm}. Even though this approach results in a high compute time compared to other existing techniques, it allows to compare the different LLMs in terms of stability and robustness with simplified generations. To provide an overview, in~\cite{wang23-valle} the codes are predicted using an autoregressive model for the first codebook and a non-autoregressive model for the residuals, which would require two separate models for each configuration proposed. In~\cite{borsos23-soundstorm}, it is proposed to use MaskGIT~\citep{chang22-maskgit} to predict the codebook-time matrix, but that results in a non-streamable model as the codes are generated non-causally. A new more advanced technique is proposed in~\cite{copet23-musicgen}, called the \textit{delay pattern}, but its implementation requires multiple heads and a latent space of higher dimensionality per model.
\subsection{Inference}
At inference time, the LLM generates the discrete non-speaker tokens and they are converted into quantized embeddings using the RVQ codebooks. Then, the speaker is defined by computing the speaker embedding from a reference utterance and concatenating that speaker embedding with the generated quantized non-speaker features. The result is fed to BigVGAN to create the speech with the target identity. As most of the constant speech information in the reference, such as mean speaker identity, is encoded in the self-supervised speaker embedding which is fed to the decoder, the LLM should be able to generate the discrete non-speaker tokens based only on the text to generate.

Based on this intuition, two types of prompting are defined: 1)~Text-prompting, where the text to generate is fed to the LLM after encoding it. If available, a reference embedding can also be fed to provide some guidance during the generation, e.g. from a trainable reference encoder. 2)~Speech-prompting, where in addition to the text-prompting inputs, the transcription of the reference and the discrete speaker-disentangled tokens corresponding to that transcription are fed to the LLM. This approach is already used in SOTA models such as VALL-E~\citep{wang23-valle}, SPEAR-TTS~\citep{kharitonov23-speartts} and SoundStorm~\citep{borsos23-soundstorm}.
\subsection{Applications}

The main purpose of this system is to be used for voice conversion and text-to-speech, so those are the use cases evaluated in this work. Nevertheless, it is important to notice that predicting speaker-disentangled codes with the LLM and training with high-quality data allow for other applications by changing the prompting: 1) Style transfer, by conditioning the LLM with speech codes of a style reference and the decoder with a speaker embedding of a speaker ID reference. 2) Super-resolution and denoising, as WavLM takes 16 kHz input waveforms and the model generates 24 kHz clean waveforms. Speech-prompting can be done to generate the same text of the reference, and as confirmed by the results shown later in Section~\ref{subsec:comparison-yourtts}, the signal quality of the original recordings could be improved by the synthetic speech.

\section{Metrics}
The models trained in this work are compared in several dimensions using objective and subjective metrics.

\textbf{Content stability, correctness and intelligibility.} The intelligibility and correctness of the generated speech is important to detect instabilities in the content generation. These instabilities can be due to repetitions, hallucinations and other mistakes during the sampling of the speech. To quantify the overall performance of each model in this dimension, the word error rate (WER) is used, which will be lower the more stable the LLM is. To transcribe the audios for the WER computation, Whisper~\citep{radford22-whisper} is used.

\textbf{Speaker similarity.} The speaker similarity to the target speaker determines the capacity of the model to clone its pitch given a reference recording. In case of using speaker-disentangled codes, this speaker similarity is mostly determined by the decoder, although it could vary depending on the prosody generated by the LLM. To facilitate reproducibility, the objective metric speaker embedding cosine similarity (SECS) using WavLM-TDNN is going to be used to measure speaker similarity, as done by~\cite{wang23-valle, le23-voicebox}. In the particular case of the voice conversion models, the speaker similarity is measured also using a MUSHRA test given the central importance of the speaker similarity in the voice conversion task, which is abbreviated as MUSHRA-S.

\textbf{Source prosody fidelity in voice conversion.} The F0 correlation between the converted and the source utterances can be used as a proxy of how well the voice conversion model is keeping the source prosody. As the SSVC model is also used to generate speech in an LLM-based TTS system, the F0 correlation is important to ensure the model follows the prosody of the discrete non-speaker codes predicted by the LLM.

\textbf{Signal quality and naturalness.} The signal quality and naturalness of the generated speech, both in VC and TTS, are measured using MUSHRA tests, abbreviated as MUSHRA-Q and MUSHRA-N respectively. These evaluations are conducted with 100 testers and 10 submissions per tester. To determine if there are significant differences between the MUSHRA scores of two systems with a 95\% confidence, paired t-tests must provide a p-value under 0.05. The best models, after performing the paired t-tests, are highlighted in bold in the tables.

\section{Experiments}
\subsection{Setup} \label{subsec:exp-setup}
\textbf{Data.} All the models presented have been trained at 24 kHz over public domain speech data and an internal subset of 3k hours of proprietary data comprised of high-quality speech. All the noisy and low-quality samples are filtered as part of the data preprocessing, leaving 44k total hours of clean and high-quality speech. Utterances longer than 16.25 seconds are not considered during training to avoid out-of-memory errors.

Regarding the evaluations, SSVC is tested using unseen source and target speakers to measure the robustness of the disentanglement of its codes over out-of-distribution data. The TTS systems are compared over unseen content and unseen speakers to test their stability at zero-shot generation and measure their robustness with out-of-distribution data. These unseen content and speakers come from LibriTTS~\citep{zen19-libritts} test-other and LibriTTS test-clean datasets respectively. In the case of the TTS evaluations, the references from test-clean are selected to be between 3 and 4 seconds long.

\textbf{Baselines.} We consider three baselines, one for the self-supervised voice conversion model and two for the LLM-based TTS: 1)~FreeVC~\citep{li22-freevc}, a SOTA text-free VC model based on WavLM that learns to disentangle content and speaker using parallel data. 2)~An LLM trained without reference embedding and with SOTA entangled codes extracted from EnCodec~\citep{défossez22-encodec}. EnCodec has already been used in SOTA TTS systems like VALL-E~\citep{wang23-valle}, VALL-E X~\citep{zhang23-vallex} and Bark\footnote{\url{https://github.com/suno-ai/bark}.}.  The selected configuration of EnCodec is 4 codes per time step, with 1024 codes per codebook, to have a similar sequence length than when using the self-supervised speaker-disentangled codes. With the downsampling ratio of EnCodec, there are 300 codes per second. 3)~YourTTS, following the procedure of VALL-E and VoiceBox~\citep{le23-voicebox}, as both models are not publicly available to directly compare against them. Note that while we use the pre-trained checkpoints of FreeVC and YoursTTS from Coqui AI\footnote{\url{https://github.com/coqui-ai/TTS}}, we trained the LLM with pre-trained EnCodec features baseline with the same training data that we used to train the models described below.

\textbf{Models.} In this work, three models have been trained using self-supervised features: a self-supervised voice conversion model, as proposed in Section~\ref{subsec:ssl-vc}, and two LLMs with the same configuration as proposed by~\cite{wang23-valle}, i.e., transformer decoders with 12 layers, $\mathbb{R}^{1024}$ model dimension, $\mathbb{R}^{4096}$ feed-forward dimension and 16 attention heads. The SSVC model has been trained with $\lambda_0 = 1/301$ and $\lambda_1 = \lambda_2 = \lambda_3 = 100/301$, as those values provided the best results in terms of speaker disentanglement, compensating the reconstruction loss of BigVGAN. Also, the purpose of training two separate LLMs with speaker-disentangled codes is measuring the stability improvement or degradation when a reference embedding is introduced to provide more context to the model.

Consequently, the first LLM is trained without a reference embedding and with speaker-disentangled self-supervised codes, and it is abbreviated as LSSL-NR. The RVQ layer is composed of 4 codeboks, each 512 codes. The BigVGAN model of the decoder has 6 upsampling layers, with a total upsampling ratio of 1:480 as the sampling frequency of the input waveforms expected by WavLM is 16 kHz, its downsampling ratio is 320:1 and the sampling frequency of the output signal is 24 kHz. It can then be easily proved that with this configuration there are 200 codes per second, so SSVC has less temporal resolution than EnCodec.

Then, a second LLM is trained with a reference encoder and with speaker-disentangled self-supervised codes, abbreviated as LSSL-R. The configuration of SSVC for this model is the same as for LSSL-NR. The reference encoder takes the spectrogram of the reference as input and combines it with convolutional layers and a self-attention layer to output a final embedding representing its content. During training, the same utterance is provided to the reference encoder and SSVC, and at inference the LLM has to generate a content different to that of the reference.

All the LLMs take as input high- and low-level embeddings of the text, separated by a learnable embedding. The high-level embeddings are extracted from a pre-trained BERT model, to provide a semantic sense to the LLM as done in Bark, and the low-level embeddings are learnable embeddings after tokenizing the text with the BPE tokenizer used by~\cite{radford19-gpt2}.

\textbf{Training.} The self-supervised voice conversion model has been trained with a batch size of 16 utterances for 950K steps using AdamW, with a constant learning rate of 1e-4, betas 0.8 and 0.99 and weight decay 0.01. Regarding the LLMs, the largest possible batch size, constrained by the 16GB of GPU memory, has been selected. In the case of EnCodec, this results in a batch size of 2.400 codes, and in the case of the self-supervised audio codec in a batch size of 3.200 codes. The LLMs are trained using a scheduler with a linear warmup of 10.000 steps, until the maximum learning rate of 5e-4 is reached, and then it is decreased using a cosine decay to 10\% of its maximum value, following~\cite{brown20-gpt3, touvron23-llama2}. All the models have been trained around 120K steps, although the baseline has been trained slightly longer to compensate the lower batch size.

\subsection{Comparison to SOTA VC model in unseen to unseen setup}
Table~\ref{tab:vc-comparison} shows SSVC achieves a lower WER, a higher F0 correlation to the source utterance and a higher speaker similarity than FreeVC, both with SECS and subjective evaluations. Consequently, the self-supervised approach based on contrastive learning is more robust to unseen speakers than the supervised approach trained with parallel data. Furthermore, the self-supervised approach has a lower training cost and a higher accessibility as it only requires raw audio, which can be collected from many multiple sources and is available even for uncommon languages.

\begin{table*}[t]
\caption{Comparison of SSVC model against FreeVC.}
\label{tab:vc-comparison}
\vskip 0.15in
\begin{center}
\begin{small}
\begin{sc}
\begin{tabular}{l|llll}
\multicolumn{1}{c}{\bf Model}  &\multicolumn{1}{c}{\bf WER} &\multicolumn{1}{c}{\bf SECS} &\multicolumn{1}{c}{\bf F0 corr. to source} &\multicolumn{1}{c}{\bf MUSHRA-S} \\ \hline
Source speaker recordings &5.7 &0.644 &n/a &n/a \\ 
Target speaker recordings &n/a &0.949 &0.069 &77.4 \\ \hline
FreeVC         &7.4 &0.814 &0.538 &72.2 \\
SSVC  &\textbf{6.0} &\textbf{0.835} &\textbf{0.593} &\textbf{74.0}
\end{tabular}
\end{sc}
\end{small}
\end{center}
\vskip -0.1in
\end{table*}

\subsection{Comparison of zero-shot TTS models based on speaker-entangled and disentangled codes}
Table~\ref{tab:llm-results} shows the performance of each LLM-based TTS model in terms of WER and SECS, with the two possible prompting methods discussed. From the results, there are several conclusions that can be drawn: 1) Text-prompting shows a significantly higher content stability than speech-prompting, as the WER is lower in all models, independently of the codes used or the presence of a reference embedding. 2) Using a reference embedding slightly improves the speaker similarity when performing text-prompting, but the content stability is greatly degraded with a 16.5pp drop in the WER. Furthermore, the model shows a significantly higher naturalness when no reference embedding is provided. 3) The baseline cannot be used in practice with text-prompting as the speaker is not cloned, with an SECS metric close to the random speaker pairing of the original recordings, due to the speaker being randomly generated by the LLM based on the conditioning text. Thus, the stability provided by text-prompting can only be leveraged by using speaker-disentangled codes, as the speaker identity is then determined in the decoder. 4) Self-supervised speaker-disentangled codes generalize better to unseen speakers, as can be seen by the significantly higher WER drop with EnCodec representations versus the WER drop obtained with the self-supervised speech representations at speech-prompting, 34.6pp and 4.3pp respectively.

\begin{table*}[t]
\caption{LLM-based TTS systems model comparison.}
\label{tab:llm-results}
\vskip 0.15in
\begin{center}
\begin{small}
\begin{sc}
\begin{tabular}{l|lll|lll}
\multicolumn{1}{c}{} &\multicolumn{3}{c}{\bf Text-prompting} &\multicolumn{3}{c}{\bf Speech-prompting} \\
\multicolumn{1}{c}{\bf Model}  &\multicolumn{1}{c}{\bf WER} &\multicolumn{1}{c}{\bf SECS} &\multicolumn{1}{c}{\bf MUSHRA-N} &\multicolumn{1}{c}{\bf WER} &\multicolumn{1}{c}{\bf SECS} &\multicolumn{1}{c}{\bf MUSHRA-N} \\ \hline
Original recordings (GT) &5.7 &0.644 &74.0 &5.7 &0.644 &76.3 \\ \hline
LLM with EnCodec  &10.1 &0.681 &68.6 &44.7 &0.881 &71.0 \\
LSSL-NR             &\textbf{4.7} &0.897 &\textbf{75.5} &\textbf{9.0} &\textbf{0.928} &\textbf{78.1} \\
LSSL-R             &21.2 &\textbf{0.909} &68.3 &23.0 &0.905 &76.1 \\
\end{tabular}
\end{sc}
\end{small}
\end{center}
\vskip -0.1in
\end{table*}

\subsection{Comparison to SOTA TTS model in zero-shot setup} \label{subsec:comparison-yourtts}
Table~\ref{tab:yourtts-results} shows the comparison in terms of intelligibilty, speaker similarity, quality and naturalness with the SOTA TTS model YourTTS~\citep{casanova22-yourtts}, using the best model both in terms of stability and naturalness, LSSL-NR. The results show that LSSL-NR with text-prompting surpasses the current SOTA in WER, quality and naturalness. Furthermore, the naturalness of the generations of LSSL-NR is not significantly different than that of the original recordings and the quality is significantly higher, which explains why LSSL-NR achieves a lower WER than the ground truth. The original recordings are part of the LibriTTS test-other subset, which is why the generated samples have a higher quality as the models have been trained only over high-quality data.

\begin{table*}[t]
\caption{Comparison with SOTA model YourTTS.}
\label{tab:yourtts-results}
\vskip 0.15in
\begin{center}
\begin{small}
\begin{sc}
\begin{tabular}{l|llll}
\multicolumn{1}{c}{\bf Model}  &\multicolumn{1}{c}{\bf WER} &\multicolumn{1}{c}{\bf SECS} &\multicolumn{1}{c}{\bf MUSHRA-Q} &\multicolumn{1}{c}{\bf MUSHRA-N} \\ \hline
Original recordings (GT) &5.7 &0.644 &71.8 &\textbf{70.6} \\ \hline
YourTTS         &8.4 &\textbf{0.932} &59.6 &50.8 \\
LSSL-NR (text-prompting)             &\textbf{4.7} &0.897 &\textbf{75.5} &\textbf{70.7} \\
LSSL-NR (speech-prompting)             &9.0 &0.928 &\textbf{75.6} &\textbf{71.2} \\
\end{tabular}
\end{sc}
\end{small}
\end{center}
\vskip -0.1in
\end{table*}

\section{Conclusion and Discussion}
This paper presents a new architecture to learn speaker-disentangled self-supervised representations directly from waveforms, named SSVC. The experiments show that this model outperforms current SOTA voice conversion systems when evaluated over unseen speakers, improving the naturalness, intelligibility and speaker similarity of the converted waveforms. This robustness with out-of-distribution data comes from learning the speaker identity using contrastive learning instead of supervised learning, and not being dependent on parallel data and its quality. Self-supervised learning allows the model to learn high level and generalizable representations from the underlying data as already proved by~\cite{schneider19-wav2vec}, which in this work results in robust speaker embeddings and quantized speaker-disentangled representations.

Then, LLMs are trained to predict these speaker-disentangled tokens, allowing to perform zero-shot TTS conditioning the LLMs only with the text to be generated as the speaker embedding is fed to the decoder. In previous SOTA models such as VALL-E, SPEAR-TTS and SoundStorm the speaker cannot not be cloned conditioning only with text, due to using entangled representations which result in random speakers. This type of conditioning, called text-prompting in this work, has also been proven to be the most stable in terms of hallucinations and repetitions with the lowest WER. Furthermore, it achieves the highest naturalness in all cases statistically significant and the highest speaker similarity of all LLM models when used with the speech-prompting technique.

\textbf{Limitations.} Using speaker-disentangled self-supervised representations enables to control the speaker in the decoder and allows the LLM focus mostly on the content and prosody. Nevertheless, there is no control over the prosody unless speech-prompting or a reference embedding conditioning is used, which results in more instabilities during the generation and a higher WER. This trade-off between stability and prosody control might come from a natural skew from the model towards predicting ``safe'', highly probable, prosody cuntours given the context.

Additionally, the sampling frequency of SSVC is 200 codes per second, so it has high compute time at inference which complicates its usage in speech streaming applications. This could be addressed in future research by applying downsampling layers to the SSL features before quantizing them or by predicting the RVQ codes using more advanced techniques like the delay pattern introduced by~\cite{copet23-musicgen}.

\textbf{Broader impact.} A self-supervised voice conversion system, that can be trained with non-parallel data and no additional annotations, can help democratize the access to voice conversion systems and its training over underrepresented languages. Additionally, given the improved stability and zero-shot speaker cloning capabilities of LLM-based TTS models trained over these speaker-disentangled representations, these reliable and expressive zero-shot TTS systems could be applied to help people with disabilities communicate and to create more natural human-computer interfaces.

\section*{Ethics Statement}
TTS and VC models with zero-shot voice cloning capabilities can be truly helpful to improve current user experiences, particularly in human-computer interfaces. However, these technologies could also be used for unethical purposes, such as deepfakes or identity theft. The models presented in this work are meant only for the first applications, and any other use is completely disapproved by the authors.

\bibliography{references}

\begin{thebibliography}{38}
\providecommand{\natexlab}[1]{#1}
\providecommand{\url}[1]{\texttt{#1}}
\expandafter\ifx\csname urlstyle\endcsname\relax
  \providecommand{\doi}[1]{doi: #1}\else
  \providecommand{\doi}{doi: \begingroup \urlstyle{rm}\Url}\fi

\bibitem[Betker(2023)]{betker23-tortoise}
Betker, J.
\newblock Better speech synthesis through scaling.
\newblock \emph{arXiv preprint arXiv:2305.07243}, 2023.

\bibitem[Borsos et~al.(2023{\natexlab{a}})Borsos, Marinier, Vincent,
  Kharitonov, Pietquin, Sharifi, Roblek, Teboul, Grangier, Tagliasacchi,
  et~al.]{borsos23-audiolm}
Borsos, Z., Marinier, R., Vincent, D., Kharitonov, E., Pietquin, O., Sharifi,
  M., Roblek, D., Teboul, O., Grangier, D., Tagliasacchi, M., et~al.
\newblock Audiolm: a language modeling approach to audio generation.
\newblock \emph{IEEE/ACM Transactions on Audio, Speech, and Language
  Processing}, 2023{\natexlab{a}}.

\bibitem[Borsos et~al.(2023{\natexlab{b}})Borsos, Sharifi, Vincent, Kharitonov,
  Zeghidour, and Tagliasacchi]{borsos23-soundstorm}
Borsos, Z., Sharifi, M., Vincent, D., Kharitonov, E., Zeghidour, N., and
  Tagliasacchi, M.
\newblock Soundstorm: Efficient parallel audio generation.
\newblock \emph{arXiv preprint arXiv:2305.09636}, 2023{\natexlab{b}}.

\bibitem[Brown et~al.(2020)Brown, Mann, Ryder, Subbiah, Kaplan, Dhariwal,
  Neelakantan, Shyam, Sastry, Askell, et~al.]{brown20-gpt3}
Brown, T., Mann, B., Ryder, N., Subbiah, M., Kaplan, J.~D., Dhariwal, P.,
  Neelakantan, A., Shyam, P., Sastry, G., Askell, A., et~al.
\newblock Language models are few-shot learners.
\newblock \emph{Advances in neural information processing systems},
  33:\penalty0 1877--1901, 2020.

\bibitem[Casanova et~al.(2022)Casanova, Weber, Shulby, Junior, G{\"o}lge, and
  Ponti]{casanova22-yourtts}
Casanova, E., Weber, J., Shulby, C.~D., Junior, A.~C., G{\"o}lge, E., and
  Ponti, M.~A.
\newblock Your{TTS}: Towards zero-shot multi-speaker tts and zero-shot voice
  conversion for everyone.
\newblock In \emph{International Conference on Machine Learning}, pp.\
  2709--2720. PMLR, 2022.

\bibitem[Chang et~al.(2022)Chang, Zhang, Jiang, Liu, and
  Freeman]{chang22-maskgit}
Chang, H., Zhang, H., Jiang, L., Liu, C., and Freeman, W.~T.
\newblock Maskgit: Masked generative image transformer.
\newblock In \emph{Proceedings of the IEEE/CVF Conference on Computer Vision
  and Pattern Recognition}, pp.\  11315--11325, 2022.

\bibitem[Chen et~al.(2022{\natexlab{a}})Chen, Xu, and
  Yu]{chen22-dataaug-nonparallelvc}
Chen, B., Xu, Z., and Yu, K.
\newblock Data augmentation based non-parallel voice conversion with
  frame-level speaker disentangler.
\newblock \emph{Speech Communication}, 136:\penalty0 14--22,
  2022{\natexlab{a}}.

\bibitem[Chen et~al.(2022{\natexlab{b}})Chen, Wang, Chen, Wu, Liu, Chen, Li,
  Kanda, Yoshioka, Xiao, et~al.]{chen22-wavlm}
Chen, S., Wang, C., Chen, Z., Wu, Y., Liu, S., Chen, Z., Li, J., Kanda, N.,
  Yoshioka, T., Xiao, X., et~al.
\newblock Wavlm: Large-scale self-supervised pre-training for full stack speech
  processing.
\newblock \emph{IEEE Journal of Selected Topics in Signal Processing},
  16\penalty0 (6):\penalty0 1505--1518, 2022{\natexlab{b}}.

\bibitem[Chung et~al.(2021)Chung, Zhang, Han, Chiu, Qin, Pang, and
  Wu]{chung2021-w2vbert}
Chung, Y.-A., Zhang, Y., Han, W., Chiu, C.-C., Qin, J., Pang, R., and Wu, Y.
\newblock W2v-bert: Combining contrastive learning and masked language modeling
  for self-supervised speech pre-training.
\newblock In \emph{2021 IEEE Automatic Speech Recognition and Understanding
  Workshop (ASRU)}, pp.\  244--250. IEEE, 2021.

\bibitem[Copet et~al.(2023)Copet, Kreuk, Gat, Remez, Kant, Synnaeve, Adi, and
  D{\'e}fossez]{copet23-musicgen}
Copet, J., Kreuk, F., Gat, I., Remez, T., Kant, D., Synnaeve, G., Adi, Y., and
  D{\'e}fossez, A.
\newblock Simple and controllable music generation.
\newblock \emph{arXiv preprint arXiv:2306.05284}, 2023.

\bibitem[D{\'e}fossez et~al.(2022)D{\'e}fossez, Copet, Synnaeve, and
  Adi]{défossez22-encodec}
D{\'e}fossez, A., Copet, J., Synnaeve, G., and Adi, Y.
\newblock High fidelity neural audio compression.
\newblock \emph{arXiv preprint arXiv:2210.13438}, 2022.

\bibitem[Devlin et~al.(2019)Devlin, Chang, Lee, and Toutanova]{devlin19-bert}
Devlin, J., Chang, M.-W., Lee, K., and Toutanova, K.
\newblock {BERT}: Pre-training of deep bidirectional transformers for language
  understanding.
\newblock In \emph{Conference of the North {A}merican Chapter of the
  Association for Computational Linguistics: Human Language Technologies},
  2019.

\bibitem[Elizalde et~al.(2023)Elizalde, Deshmukh, Al~Ismail, and
  Wang]{elizalde22-clap}
Elizalde, B., Deshmukh, S., Al~Ismail, M., and Wang, H.
\newblock Clap learning audio concepts from natural language supervision.
\newblock In \emph{ICASSP 2023-2023 IEEE International Conference on Acoustics,
  Speech and Signal Processing (ICASSP)}, pp.\  1--5. IEEE, 2023.

\bibitem[gil Lee et~al.(2023)gil Lee, Ping, Ginsburg, Catanzaro, and
  Yoon]{lee2023-bigvgan}
gil Lee, S., Ping, W., Ginsburg, B., Catanzaro, B., and Yoon, S.
\newblock Big{VGAN}: A universal neural vocoder with large-scale training.
\newblock In \emph{International Conference on Learning Representations}, 2023.

\bibitem[Hsu et~al.(2021)Hsu, Bolte, Tsai, Lakhotia, Salakhutdinov, and
  Mohamed]{hsu21-hubert}
Hsu, W.-N., Bolte, B., Tsai, Y.-H.~H., Lakhotia, K., Salakhutdinov, R., and
  Mohamed, A.
\newblock Hubert: Self-supervised speech representation learning by masked
  prediction of hidden units.
\newblock \emph{IEEE/ACM Transactions on Audio, Speech, and Language
  Processing}, 29:\penalty0 3451--3460, 2021.

\bibitem[Huang et~al.(2022)Huang, Yang, Hayashi, and Toda]{huang22-s3rvc}
Huang, W.-C., Yang, S.-W., Hayashi, T., and Toda, T.
\newblock A comparative study of self-supervised speech representation based
  voice conversion.
\newblock \emph{IEEE Journal of Selected Topics in Signal Processing},
  16\penalty0 (6):\penalty0 1308--1318, 2022.

\bibitem[Kameoka et~al.(2018)Kameoka, Kaneko, Tanaka, and
  Hojo]{kameoka18-starganvc}
Kameoka, H., Kaneko, T., Tanaka, K., and Hojo, N.
\newblock Stargan-vc: Non-parallel many-to-many voice conversion using star
  generative adversarial networks.
\newblock In \emph{2018 IEEE Spoken Language Technology Workshop (SLT)}, pp.\
  266--273. IEEE, 2018.

\bibitem[Kaneko et~al.(2019)Kaneko, Kameoka, Tanaka, and
  Hojo]{kaneko19-cycleganvc2}
Kaneko, T., Kameoka, H., Tanaka, K., and Hojo, N.
\newblock Cycle{GAN}-vc2: Improved cyclegan-based non-parallel voice
  conversion.
\newblock In \emph{ICASSP 2019-2019 IEEE International Conference on Acoustics,
  Speech and Signal Processing (ICASSP)}, pp.\  6820--6824. IEEE, 2019.

\bibitem[Kharitonov et~al.(2023)Kharitonov, Vincent, Borsos, Marinier, Girgin,
  Pietquin, Sharifi, Tagliasacchi, and Zeghidour]{kharitonov23-speartts}
Kharitonov, E., Vincent, D., Borsos, Z., Marinier, R., Girgin, S., Pietquin,
  O., Sharifi, M., Tagliasacchi, M., and Zeghidour, N.
\newblock Speak, read and prompt: High-fidelity text-to-speech with minimal
  supervision.
\newblock \emph{arXiv preprint arXiv:2302.03540}, 2023.

\bibitem[Le et~al.(2023)Le, Vyas, Shi, Karrer, Sari, Moritz, Williamson,
  Manohar, Adi, Mahadeokar, et~al.]{le23-voicebox}
Le, M., Vyas, A., Shi, B., Karrer, B., Sari, L., Moritz, R., Williamson, M.,
  Manohar, V., Adi, Y., Mahadeokar, J., et~al.
\newblock Voicebox: Text-guided multilingual universal speech generation at
  scale.
\newblock \emph{arXiv preprint arXiv:2306.15687}, 2023.

\bibitem[Li et~al.(2023)Li, Tu, and Xiao]{li22-freevc}
Li, J., Tu, W., and Xiao, L.
\newblock Free{VC}: Towards high-quality text-free one-shot voice conversion.
\newblock In \emph{ICASSP 2023-2023 IEEE International Conference on Acoustics,
  Speech and Signal Processing (ICASSP)}, pp.\  1--5. IEEE, 2023.

\bibitem[Lin et~al.(2021)Lin, Chien, Lin, Lee, and Lee]{lin21-fragmentvc}
Lin, Y.~Y., Chien, C.-M., Lin, J.-H., Lee, H.-y., and Lee, L.-s.
\newblock Fragment{VC}: Any-to-any voice conversion by end-to-end extracting
  and fusing fine-grained voice fragments with attention.
\newblock In \emph{ICASSP 2021-2021 IEEE International Conference on Acoustics,
  Speech and Signal Processing (ICASSP)}, pp.\  5939--5943. IEEE, 2021.

\bibitem[Merritt et~al.(2022)Merritt, Ezzerg, Bili{\'n}ski, Proszewska, Pokora,
  Barra-Chicote, and Korzekwa]{merritt2022-textfree-nfvc}
Merritt, T., Ezzerg, A., Bili{\'n}ski, P., Proszewska, M., Pokora, K.,
  Barra-Chicote, R., and Korzekwa, D.
\newblock Text-free non-parallel many-to-many voice conversion using
  normalising flow.
\newblock In \emph{ICASSP 2022-2022 IEEE International Conference on Acoustics,
  Speech and Signal Processing (ICASSP)}, pp.\  6782--6786. IEEE, 2022.

\bibitem[Qian et~al.(2019)Qian, Zhang, Chang, Yang, and
  Hasegawa-Johnson]{qian19-autovc}
Qian, K., Zhang, Y., Chang, S., Yang, X., and Hasegawa-Johnson, M.
\newblock Auto{VC}: Zero-shot voice style transfer with only autoencoder loss.
\newblock In \emph{International Conference on Machine Learning}, pp.\
  5210--5219. PMLR, 2019.

\bibitem[Radford et~al.(2019)Radford, Wu, Child, Luan, Amodei, Sutskever,
  et~al.]{radford19-gpt2}
Radford, A., Wu, J., Child, R., Luan, D., Amodei, D., Sutskever, I., et~al.
\newblock Language models are unsupervised multitask learners.
\newblock \emph{OpenAI blog}, 1\penalty0 (8):\penalty0 9, 2019.

\bibitem[Radford et~al.(2021)Radford, Kim, Hallacy, Ramesh, Goh, Agarwal,
  Sastry, Askell, Mishkin, Clark, et~al.]{radford21-clip}
Radford, A., Kim, J.~W., Hallacy, C., Ramesh, A., Goh, G., Agarwal, S., Sastry,
  G., Askell, A., Mishkin, P., Clark, J., et~al.
\newblock Learning transferable visual models from natural language
  supervision.
\newblock In \emph{International conference on machine learning}, pp.\
  8748--8763. PMLR, 2021.

\bibitem[Radford et~al.(2023)Radford, Kim, Xu, Brockman, McLeavey, and
  Sutskever]{radford22-whisper}
Radford, A., Kim, J.~W., Xu, T., Brockman, G., McLeavey, C., and Sutskever, I.
\newblock Robust speech recognition via large-scale weak supervision.
\newblock In \emph{International Conference on Machine Learning}, pp.\
  28492--28518. PMLR, 2023.

\bibitem[Saito et~al.(2018)Saito, Ijima, Nishida, and Takamichi]{saito18-vae}
Saito, Y., Ijima, Y., Nishida, K., and Takamichi, S.
\newblock Non-parallel voice conversion using variational autoencoders
  conditioned by phonetic posteriorgrams and d-vectors.
\newblock In \emph{2018 IEEE International Conference on Acoustics, Speech and
  Signal Processing (ICASSP)}, pp.\  5274--5278. IEEE, 2018.

\bibitem[Schneider et~al.(2019)Schneider, Baevski, Collobert, and
  Auli]{schneider19-wav2vec}
Schneider, S., Baevski, A., Collobert, R., and Auli, M.
\newblock wav2vec: Unsupervised pre-training for speech recognition.
\newblock \emph{arXiv preprint arXiv:1904.05862}, 2019.

\bibitem[Schnell et~al.(2021)Schnell, Huybrechts, Perz, Drugman, and
  Lorenzo-Trueba]{schnell21-emocat}
Schnell, B., Huybrechts, G., Perz, B., Drugman, T., and Lorenzo-Trueba, J.
\newblock {EmoCat: Language-agnostic Emotional Voice Conversion}.
\newblock In \emph{ISCA Speech Synthesis Workshop (SSW 11)}, pp.\  72--77,
  2021.

\bibitem[Sisman et~al.(2020)Sisman, Yamagishi, King, and
  Li]{sisman20-vcoverview}
Sisman, B., Yamagishi, J., King, S., and Li, H.
\newblock An overview of voice conversion and its challenges: From statistical
  modeling to deep learning.
\newblock \emph{IEEE/ACM Transactions on Audio, Speech, and Language
  Processing}, 29:\penalty0 132--157, 2020.

\bibitem[Touvron et~al.(2023)Touvron, Martin, Stone, Albert, Almahairi, Babaei,
  Bashlykov, Batra, Bhargava, Bhosale, et~al.]{touvron23-llama2}
Touvron, H., Martin, L., Stone, K., Albert, P., Almahairi, A., Babaei, Y.,
  Bashlykov, N., Batra, S., Bhargava, P., Bhosale, S., et~al.
\newblock Llama 2: Open foundation and fine-tuned chat models.
\newblock \emph{arXiv preprint arXiv:2307.09288}, 2023.

\bibitem[Wang et~al.(2023)Wang, Chen, Wu, Zhang, Zhou, Liu, Chen, Liu, Wang,
  Li, et~al.]{wang23-valle}
Wang, C., Chen, S., Wu, Y., Zhang, Z., Zhou, L., Liu, S., Chen, Z., Liu, Y.,
  Wang, H., Li, J., et~al.
\newblock Neural codec language models are zero-shot text to speech
  synthesizers.
\newblock \emph{arXiv preprint arXiv:2301.02111}, 2023.

\bibitem[wen Yang et~al.(2021)wen Yang, Chi, Chuang, Lai, Lakhotia, Lin, Liu,
  Shi, Chang, Lin, hsien Huang, Tseng, tik Lee, Liu, Huang, Dong, Li, Watanabe,
  rahman Mohamed, and yi~Lee]{yang2021-superb}
wen Yang, S., Chi, P.-H., Chuang, Y.-S., Lai, C.-I., Lakhotia, K., Lin, Y.~Y.,
  Liu, A.~T., Shi, J., Chang, X., Lin, G.-T., hsien Huang, T., Tseng, W.-C.,
  tik Lee, K., Liu, D.-R., Huang, Z., Dong, S., Li, S.-W., Watanabe, S., rahman
  Mohamed, A., and yi~Lee, H.
\newblock {SUPERB}: Speech processing universal performance benchmark.
\newblock In \emph{Interspeech}, 2021.

\bibitem[Yang et~al.(2023)Yang, Liu, Huang, Tian, Weng, and
  Zou]{yang23-hificodec}
Yang, D., Liu, S., Huang, R., Tian, J., Weng, C., and Zou, Y.
\newblock Hifi-codec: Group-residual vector quantization for high fidelity
  audio codec.
\newblock \emph{arXiv preprint arXiv:2305.02765}, 2023.

\bibitem[Zeghidour et~al.(2021)Zeghidour, Luebs, Omran, Skoglund, and
  Tagliasacchi]{zeghidour21-soundstream}
Zeghidour, N., Luebs, A., Omran, A., Skoglund, J., and Tagliasacchi, M.
\newblock Soundstream: An end-to-end neural audio codec.
\newblock \emph{IEEE/ACM Transactions on Audio, Speech, and Language
  Processing}, 30:\penalty0 495--507, 2021.

\bibitem[Zen et~al.(2019)Zen, Dang, Clark, Zhang, Weiss, Jia, Chen, and
  Wu]{zen19-libritts}
Zen, H., Dang, V., Clark, R., Zhang, Y., Weiss, R.~J., Jia, Y., Chen, Z., and
  Wu, Y.
\newblock Libri{TTS}: A corpus derived from librispeech for text-to-speech.
\newblock \emph{Interspeech 2019}, 2019.

\bibitem[Zhang et~al.(2023)Zhang, Zhou, Wang, Chen, Wu, Liu, Chen, Liu, Wang,
  Li, et~al.]{zhang23-vallex}
Zhang, Z., Zhou, L., Wang, C., Chen, S., Wu, Y., Liu, S., Chen, Z., Liu, Y.,
  Wang, H., Li, J., et~al.
\newblock Speak foreign languages with your own voice: Cross-lingual neural
  codec language modeling.
\newblock \emph{arXiv preprint arXiv:2303.03926}, 2023.

\end{thebibliography}
\bibliographystyle{icml2023}

\end{document}